\documentclass{vgtc}                          

\graphicspath{{figures/}{pictures/}{images/}{./}} 

\usepackage{times}                     

\usepackage{tabu}                      
\usepackage{booktabs}                  
\usepackage{lipsum}                    
\usepackage{mwe}                       

\usepackage{mathptmx}                  

\onlineid{0}

\vgtccategory{Research}

\vgtcinsertpkg

\renewcommand*{\backref}[1]{
}

\title{Embedding Empathy into Visual Analytics: A Framework for
Person-Centred Dementia Care}

\author{Rhiannon Owen\thanks{e-mail: r.s.owen@bangor.ac.uk}\\ %
        \scriptsize Bangor University %
\and Jonathan C. Roberts\thanks{e-mail: j.c.roberts@bangor.ac.uk}\\ %
     \scriptsize Bangor University}

\teaser{
  \centering
  \includegraphics[alt={Flow diagram with labelled steps connected by arrows, illustrating the EDVF process from initial assessment to application in dementia care.},width=\linewidth]{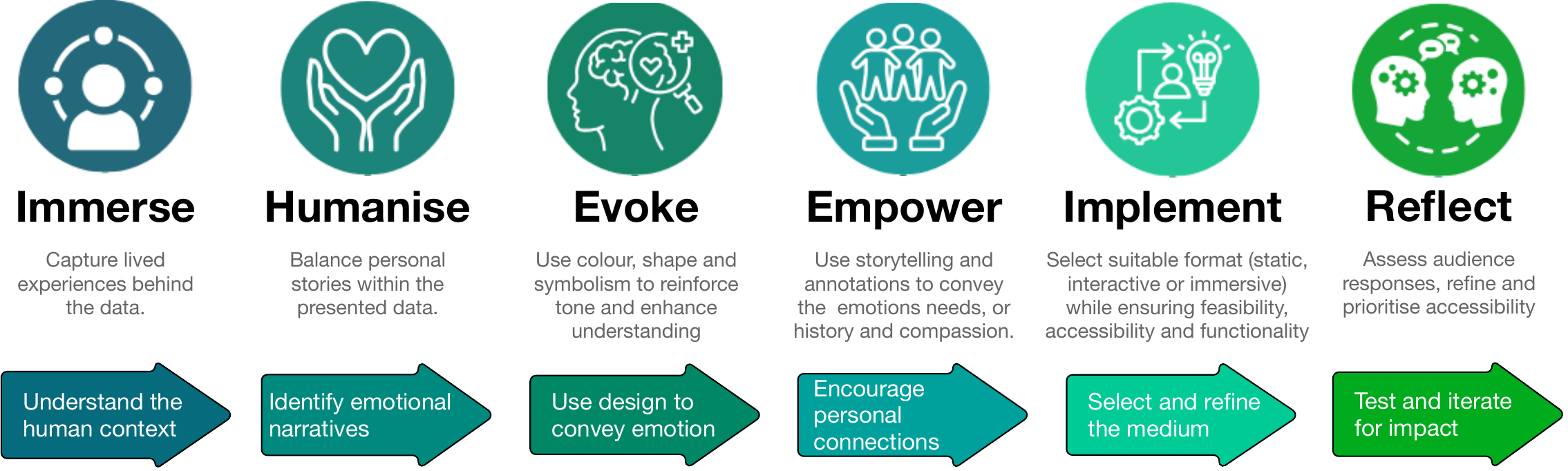}
  \caption{Diagram of the Empathy-Driven Visualisation
Framework for Dementia Care (EDVF) using arrows to show the process with brief explanations of each step.}
  \label{fig:teaser}
}

\abstract{
Dementia care requires healthcare professionals to balance a patient’s medical needs with a deep understanding of their personal needs, preferences, and emotional cues. However, current digital tools prioritise quantitative metrics over empathetic engagement, limiting caregivers' ability to develop a deeper personal understanding of their patients. This paper presents an empathy-centred visualisation framework, developed through a design study, to address this gap. The framework integrates established
principles of person-centred care with empathy mapping methodologies to encourage deeper engagement. Our methodology provides a structured approach to designing for indirect end-users, patients whose experience is shaped by a tool they may not directly interact with. To validate the framework, we conducted evaluations with
healthcare professionals, including usability testing of a working prototype and a User Experience Questionnaire (UEQ) study. Results suggest the feasibility of the framework, with participants highlighting its potential to support a more personal and empathetic relationship between medical staff and patients. The work starts to explore how empathy could be systematically embedded into visualisation design, as we contribute to ongoing efforts in the data visualisation community to support human-centred, interpretable, and ethically-aligned clinical care, addressing the urgent need to improve dementia
patients' experiences in hospital settings.
} 


\begin{document}


\firstsection{Introduction}

\maketitle

Dementia is one of the most pressing healthcare challenges globally, significantly contributing to neurological disability and mortality. Patients with dementia often face longer hospital stays and poorer health outcomes compared to those without cognitive impairments~\cite{ClissettETAL2013}. A critical contributing factor is the difficulty healthcare professionals face in establishing meaningful connections with dementia patients. Miscommunication can lead to patients being perceived as ``challenging'', which, in turn, reduces empathy and compromises the quality of care~\cite{GwernanLourida2020, baumbusch2016factors}.  

\begin{figure}[h]
    \centering
    \includegraphics[alt={Diagram showing the four stages of person-centred care: Values, Individuals, Perspective, Social.},width=1\linewidth]{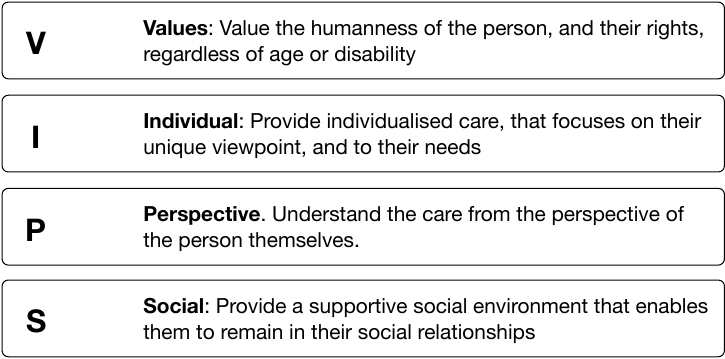}
    \caption{The VIPS framework highlighting person-centred approaches. The four core components are: Values, Individual's needs, Perspective of the service user, Supportive social environment. Diagram reproduced from Brooker and Latham~\cite{Brooker2003}.}
    \label{fig:vips-framework}
\end{figure}

To facilitate person-centred care, existing resources such as the ``This is Me'' document from the Alzheimer's Society help caregivers record personal details, such as family names, friends, pets, and current interests, particularly for individuals with dementia. However, at its core, it is a sheet of paper. Something that must be completed by the healthcare worker with the family of the patient, and can easily be misplaced or forgotten in the hustle of a busy clinical environment~\cite{banks2014dementia}. Beyond the need for accessible personal information, a greater challenge lies in fostering an empathetic connection between patients and caregivers. While individuals with dementia may not directly engage with digital tools, they should still benefit from technology. As \textbf{indirect end-users} (patients whose experience is shaped by tools they don't directly operate), they can gain significantly from innovations that prioritise their emotional and personal well-being.

Existing visualisation models, such as Seldmair's nine-stage framework~\cite{seldmair2012} and Nested model~\cite{munzner2009nested}, emphasise user-centred design but primarily focus on task efficiency rather than deep, meaningful empathy. While many visualisation studies account for user needs, they do not explicitly centre around the emotional, human experience in an accessible and actionable way. Critically, empathy in visualisation should be more than an abstract goal, it must be an integral part of the design process. This means prioritising accessibility, ensuring clarity in communication, and creating designs that accommodate diverse needs in high-stakes environments like healthcare. Visualisations should not only inform but also evoke understanding and foster deeper connections between data and the people it represents.

In contrast, our work shifts the focus toward \textit{deep empathetic engagement}, not only considering what healthcare professionals need but also embedding emotional depth that allows them to connect with patients on a more human level. To address this gap, we developed the \textbf{Empathy-Driven Visualisation Framework (EDVF)}, a structured approach to enhancing person-centred dementia care through digital means. Unlike traditional frameworks, EDVF is designed to embed deep empathy directly into the visualisation process, ensuring designs facilitate not just understanding, but human connection and empowerment. By integrating personal narratives and balancing quantitative metrics with qualitative insights, visualisations become tools of empathy that help caregivers grasp patients' realities. This humanisation allows patients to be seen as more than data, individuals with unique histories and emotional needs.

\begin{figure}[h]
    \centering
    \includegraphics[alt={Grid-based empathy map template with labelled sections for thoughts, feelings, needs, and observations, designed for mapping a dementia patient’s experience.},width=1\linewidth]{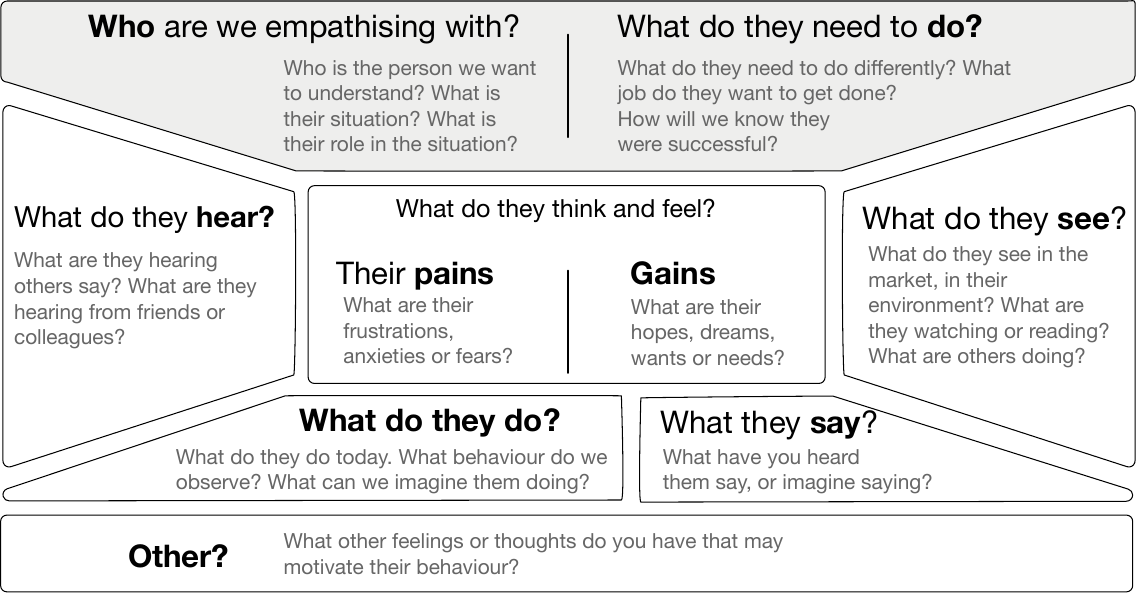}
    \caption{Empathy map canvas, reproduced by authors, from  Gray et al., \cite{GrayBrownMacanufo2010}. It could be used to map the experience and feelings of a dementia patient, highlighting emotions, concerns, and cognitive challenges.}
    \label{fig:empathy-map}
\end{figure}

The framework is adaptable and can be applied by anyone looking to incorporate empathy driven visualisation into their work, whether in healthcare, education, or other fields where human-centred design is essential. EDVF integrates the VIPS model for person-centred care~\cite{brooker2016person} with `empathy mapping' methodologies~\cite{GrayBrownMacanufo2010} (see \cref{fig:empathy-map}) to create more meaningful and intuitive interactions with patient data. The VIPS framework emphasises valuing people as individuals and understanding their perspective, enabling healthcare workers to provide a supportive environment. The method \textbf{V}alues the person, is \textbf{I}ndividual, focuses on the \textbf{P}erspective of the person, and supports them in their \textbf{S}ocial environment (see \cref{fig:vips-framework}).

Our EDVF framework is built upon six guiding principles: Immerse, Humanise, Evoke, Empower, Implement and Reflect (see \cref{fig:teaser}). These principles are realised through three core design pillars: \textit{Hierarchical Salience-Driven Design} (prioritising critical information), \textit{Task-Optimised Iconography} (enhancing clarity), and \textit{First-Person Narrative Integration} (incorporating personal perspectives to encourage empathy). This paper presents three key contributions: 
\begin{enumerate}
    \item The EDVF \textbf{structured framework for empathy-driven visualisations} applied to dementia care, systematically integrating person-centred care principles with empathetic visualisation techniques (\cref{sec:Framework})
    \item A \textbf{prototype application for healthcare professionals} demonstrating enhanced understanding and communication with dementia patients
    \item An \textbf{in-depth study involving over 30 healthcare professionals} providing qualitative insights into usability and impact (\cref{SEC:evaluation})
\end{enumerate}
By framing empathy as a structured component of healthcare design, our work contributes to the growing need for interpretable, ethical, and emotionally aware analytics in clinical environments, ultimately aiming to improve communication and health outcomes through empowered care relationships.

\section{Background}
\label{SEC:Background}

Visual analytics (VA) is becoming an essential approach in healthcare, supporting not only data-driven decision-making but also improving communication, empathy, and patient engagement~\cite{BernardIEEE, caban2015visual}. In dementia care, where patients often struggle to articulate their needs and caregivers face emotional and logistical pressures, existing tools frequently fall short, focusing on clinical efficiency while neglecting human-centred design. This research draws on over five years of embedded experience in dementia care, including site visits to hospital dementia wards, direct engagement with patients and professionals, and two studies at national healthcare events: \textbf{Study A} at the Dementia Aware Event and \textbf{Study B} at a healthcare symposium in Rhyl, UK. These efforts revealed a critical gap: while healthcare data is plentiful, few systems provide intuitive, empathetic, and actionable visualisations to support caregivers or patients.

\begin{figure*}[t]
    \centering
    \includegraphics[alt={Screenshots of the prototype interface displayed on a laptop, tablet, and phone, showing various sections, icons, and layouts.},width=\textwidth]{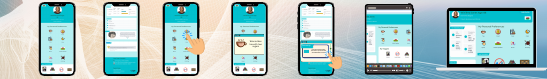}
    \caption{Prototype visualisation of the prototype and how it looks across several devices – demonstrating the prototype and the different sections it has, example icons for triggers, etc.}
    \label{fig:Prototype}
\end{figure*}

Healthcare professionals repeatedly expressed frustration with digital tools that prioritise data access but \textit{fail to convey essential patient context.} There is a clear opportunity for VA and interaction design to enhance both efficiency and empathy by humanising data, making it accessible, meaningful, and responsive to the realities of caregiving. This includes presenting information succinctly, framing it in a sensitive and inclusive way, and translating complex data into formats that are easy to understand and explain. Currently, the \textit{This is Me} document, a paper-based tool developed by the Alzheimer's Society, is the only widely used method for presenting personal information in dementia care and \textbf{no digital alternatives exist.} At \textbf{Study B}, Carol, a woman living with dementia, shared her hospital experience: \textit{``The nurses were unempathetic towards me. I felt confused and lost.''} Her distress, which ultimately led to a suicide attempt, emphasises the urgency of developing systems that empower both patients and clinicians with clarity, empathy, and support throughout the care journey. Our work addresses this challenge by developing the grounding EDVF framework, which guides the creation of clinical tools that integrate interaction design with person-centred approaches to foster empathy, improve clinician–patient communication, support personalised care, and ultimately enhance the quality of life for people living with dementia.

\section{Related Work}

This research sits at the intersection of three core domains: (1) dementia care and person-centred healthcare, (2) empathy and accessibility in visualisation, and (3) visualisation techniques for explainability and clarity.

\subsection{Dementia Care and Person-Centred Healthcare}

A hierarchical regression analysis by Zank~\cite{Zank01} revealed that the interaction of physical diseases and cognitive functioning produces a significant change in depressive symptoms and life satisfaction, dementia care presents unique challenges due to the progressive cognitive decline experienced by patients, significantly impacting communication and emotional well-being. Person-centred care (PCC), first introduced by Kitwood~\cite{kitwood1988technical}, prioritises patient dignity, identity, and emotional needs. It encourages shared decision-making between healthcare providers and patients~\cite{brooker2016person, Santana2017}. However, implementation often remains aspirational in clinical practice~\cite{Popham01032012}. As established in \cref{SEC:Background}, while respecting human dignity remains fundamental, existing tools like the paper-based \textit{This is Me} document face accessibility limitations in clinical settings. This underscores the need for visualisation systems that better integrate emotional and personal context alongside clinical data.

\subsection{Empathy and Accessibility in Visualisation}

Empathy mapping is a design thinking technique used to visually represent a person's thoughts, feelings, and needs \cite{GrayBrownMacanufo2010}, it offers a structured way to represent user emotions and challenges, supporting tools tailored to user experience. While underutilised in clinical settings, this approach has shown promise in healthcare communication~\cite{MoudatsouETAL2020, SiricharoenEmpathyMapping}. In dementia contexts, patients are often \textbf{indirect users}, yet their needs profoundly shape the tools caregivers rely on. Our approach uses empathy mapping to better inform caregiver tools by incorporating insights into patient affective states~\cite{Zank01}. Most visualisation research prioritises usability, overlooking emotional impact. Inclusive visualisation frameworks like \textit{Feminist Data Visualisation}~\cite{dignazio2020data} advocate for accessibility but don't explicitly focus on empathy. We build on this by embedding empathy as a design pillar to support emotional connection, not just comprehension.

\subsection{Visualisation for Explainability and Narrative Clarity}

Explanatory visualisation frameworks emphasise clarity and comprehension~\cite{roberts2018explanatory}, but rarely address affective engagement. Storytelling visualisation aims to improve engagement~\cite{segel2010narrative}, yet often remains data-centric. Spaulding~\cite{SpauldingDesignDriven} and Coats~\cite{CoatsPerson-CenteredNarrative} argue for co-designed, first-person narratives in patient records to foster connection approaches, our work adopts to bridge data and emotional experience. Our project introduces First-Person Narrative Integration into VA tools to elevate patient voices and support empathetic interpretation. By foregrounding the human experience within visualisations, we align with feminist and community-centred design principles~\cite{dignazio2016feminist}, recognising that care is shaped not just by individual data, but by broader cultural, clinical, and emotional contexts.
\section{The Empathetic Design Visualisation Framework}
\label{sec:Framework}

The Empathetic Design Visualisation Framework (EDVF) provides a structured methodology for embedding empathetic care and human connection into data representation. Unlike conventional approaches that prioritise efficiency or analytical clarity, EDVF foregrounds the human experience, making it especially valuable in domains such as healthcare, where data can risk dehumanising its subjects, and where emotional engagement is essential to effective care. Building on the principles outlined in \textit{``Against the dehumanisation of decision-making''}~\cite{LaDiega2018}, the framework integrates the VIPS model of person-centred dementia care, empathy mapping techniques, person-centred care (PCC) strategies, and empathetic design principles to construct a comprehensive, human-first approach to visualisation.

At its core is the concept of \textbf{immersion}, a design imperative that urges researchers to actively engage with patient communities to capture nuanced emotional contexts often absent in clinical datasets. Unlike traditional visualisation frameworks, EDVF uniquely transforms data into compassionate narratives through techniques like personalised storytelling, adaptive iconography, and colour mapping that evoke emotional tone. This helps clinicians not only interpret data but connect more deeply with the individuals behind it. However, the framework acknowledges significant challenges including ethical concerns around privacy in visualising sensitive data~\cite{correll2018ethical}, and the tendency in design processes to bypass empathy phases~\cite{zulaikhaEmpathyChallenge}. Ensuring fairness and inclusivity, particularly in global contexts, remains an ongoing responsibility~\cite{RyanFairness}.

To validate the framework, we applied it in the development of a prototype \textit{``About Me''} application. This tool was designed to challenge dehumanisation in dementia care by enabling the visual representation of patient narratives and preferences~\cite{BogmoreDemunanisation}. The application served both as a proof of concept and as an opportunity to reflect on the emotional and ethical dimensions of patient-centred visual tools. The EDVF follows a six-stage process: \textbf{Immersion}, \textbf{Humanisation}, \textbf{Evocation}, \textbf{Empowerment}, \textbf{Implementation} and \textbf{Reflection}. These stages ensure that data representations are not only informative, but also foster empathy, dignity, and emotional connection. Following the sequence in \cref{fig:teaser}, the next sections detail how each stage is implemented across various domains.

\subsection{Immersion: Developing a Deep Understanding}

Designing empathetic healthcare visualisations requires more than data analysis, it demands \textbf{immersion} in the lived experiences behind the data. This means engaging deeply with patients, clinicians, and carers to understand the complex contexts influencing health outcomes~\cite{dignazio2016feminist}. Immersion goes beyond interviews or surveys; it involves direct interaction and observation in real-world settings like volunteering in care homes to capture nuances that data alone cannot reveal. Superficial methods are insufficient for grasping deeper user requirements~\cite{zulaikhaEmpathyChallenge}. For example, spending time with patients and care staff uncovers vital personal details, such as individual preferences, triggers of distress, or comforting routines that shape care quality but may be absent from formal records. These intimate insights create a richer, empathetic understanding that respects patient dignity and individuality. Whether visualising clinical data, or patient-generated information, grounding the design in first-hand experiences ensures the visualisation reflects both the data and the human stories behind it. By bridging data and empathy, immersive engagement helps create visual tools that truly support meaningful, person-centred healthcare.

\subsection{Humanisation: Identifying the Emotional Core}

Once immersed, the next step is to \textit{humanise the data} by translating impersonal records into meaningful narratives. Humanising visualisations goes beyond storytelling; it means designing with real human needs, limitations, and experiences in mind. Data should be presented in a way that is emotionally meaningful, practical, accessible, and sensitive to the contexts of those who use it. This transformation of data into compassionate narratives helps clinicians see patients as more than data points, individuals with unique histories and emotional needs.

\begin{figure}[h]
    \centering
    \includegraphics[alt={Set of illustrated icons depicting different sensory or environmental triggers, each with distinct symbols to represent specific situations.},width=.9\columnwidth]{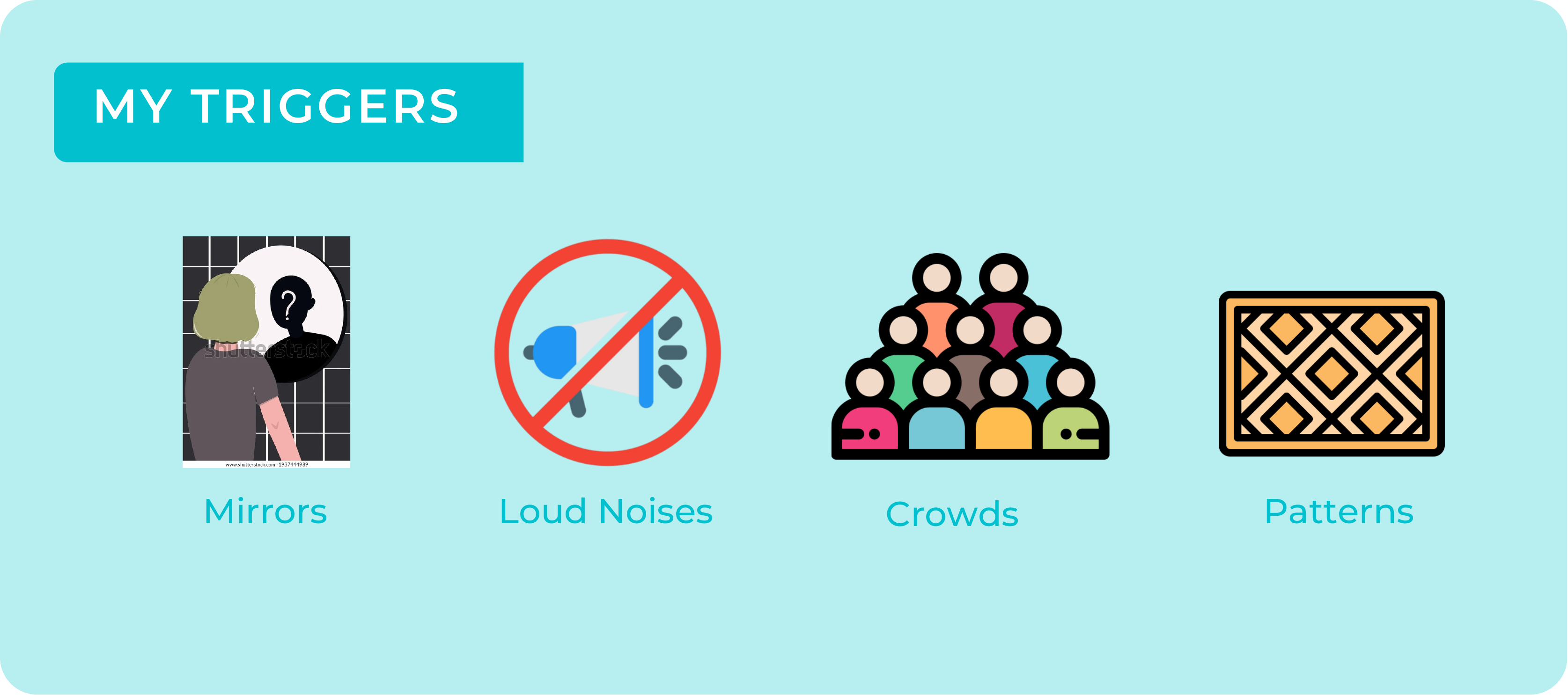}
    \caption{Icons from the prototype, representing patient-specific triggers (e.g., loud noises, crowded spaces) developed using the EDVF. These symbols allow caregivers to quickly identify and avoid distress sources while promoting empathetic understanding of individual sensitivities -- once clicked they show a reason for the trigger.}
    \vspace{-3mm}
    \label{fig:trigger-icons}
\end{figure}

In dementia care, this means moving from generic medical charts to individualised patient stories. For example, rather than simply noting \textit{“agitation triggers”}, an empathetic visualisation should reveal the emotional weight behind these moments of distress (\cref{fig:trigger-icons}). Small but significant details shape the emotional reality of patients. Even well-meaning gestures can cause unintended distress; clinical specialist Kathryn Williams at Llandudno Hospital notably explained: \textit{“I tried to calm a dementia patient by providing comfort, however, the situation became worse as she had a trigger to personal touch that I was unaware of, she really declined that day. These small things matter greatly in dementia care.”} In healthcare settings, where minor adjustments can significantly reduce anxiety and improve comfort, empathetic visualisation can turn data into clear, actionable, and emotionally resonant insights that prioritise human dignity and lived experience over abstract data points. Humanising data also involves capturing moments of comfort and familiarity, embedding personal details to support consistent, compassionate care. This aligns with research emphasising the importance of contextualised, \textit{human-centred data representation}~\cite{tversky2002animation}. Designers must consider both emotional connection and practical usability, ensuring visualisations are not only relatable but also genuinely useful.

\subsection{Evoke: Designing for Emotional Impact}

The \textbf{evocation} stage focuses on creating a visualisation that is both emotionally resonant and clear. In dementia care, this means designing visuals that trigger emotions, memories, and deeper understanding. Using elements such as colour mapping, imagery, symbolism, and storytelling helps encourage empathy and connection between caregivers, patients, and data. To encourage deeper empathetic connections between patient and carer, integrating first-person narrative elements provides valuable insights into a patient’s history, emotional triggers, and personal comforts. This approach aligns with studies suggesting that embedding patient stories and first-person narratives creates empathy and enhances person-centred care, \textit{“Our study demonstrates that a co-created patient narrative intervention provides avenues for patients and nurses to connect despite being in hectic acute care settings”}~\cite{CoatsPerson-CenteredNarrative}. A well-designed dementia care visualisation could incorporate task-optimised iconography to depict patient preferences through intuitive symbols, enabling caregivers to quickly extract key details under time constraints~\cite{annieicon, Lau2019Improving}. Highlighting the most relevant and contextually significant information allows caregivers to focus on critical details amidst overwhelming data, facilitating swift and informed decisions. For example, rather than a simple chart of agitation levels, an evocative visualisation might include personal anecdotes or calming imagery to guide caregivers in providing comfort. Ultimately, the \textbf{evoke} stage ensures data is not just seen but felt.

\subsection{Empowerment: Making the Visualisation Relatable and Actionable}

\textbf{Empowerment} ensures that healthcare professionals can meaningfully engage with the visualisation, translating data into informed, compassionate actions. In clinical environments, where time and cognitive resources are limited, clear, accessible interfaces are essential. For example, a nurse unfamiliar with a patient should quickly grasp critical care information, such as sensitivities to light or noise, without sifting through extensive notes. As highlighted by Nicole Gross~\cite{GrossEmpowerHealthcare}, digital health technologies can empower both caregivers and patients, enhancing many aspects of clinical care and helping patient-centredness~\cite{GrossEmpowerHealthcare}. However, these technologies also introduce new responsibilities for caregivers, as they navigate both empowerment and the stress of additional duties. In this way, the design of digital tools must consider the balance between providing \textit{supportive care} and not \textit{overburdening} caregivers, ensuring that empowerment translates into better outcomes, without increasing stress.

\subsection{Implementation and Reflection: Bringing the Visualisation to Life}

Successful \textbf{implementation} of healthcare visualisations requires careful attention to both technical and human factors. It involves careful consideration of how the system will benefit the \textbf{indirect end user}: the dementia patient. Therefore, designs must prioritise usability, readability (including in low-light conditions), and seamless interaction across devices~\cite{Madhavan2024}. Regular feedback loops involving patients, families, and staff support ongoing refinement, ensuring visualisations remain sensitive to patient needs and contribute to comfort and trust. Reflection is crucial: have the visualisations improved patient well-being? Do caregivers feel better informed and more connected? However, despite the critical role these visualisations play in strengthening the caregiver-patient relationship, there remains a notable gap in the literature regarding their impact on the indirect end user, the dementia patient. Given the fundamental importance of creating empathy and patient-centred care in dementia contexts, further investigation is needed to optimise visualisations that enhance both patient experience and caregiver effectiveness.

\subsection{Ethical Considerations and Limitations}
The framework acknowledges ethical challenges in visualising sensitive health data. Ensuring representations don't infringe on privacy or mislead viewers is a \textit{moral obligation}~\cite{correll2018ethical}, requiring vigilance against unethical practices even amid organisational pressures. Additionally, incorporating true fairness remains challenging as existing HCI research often neglects global contexts~\cite{RyanFairness}. These limitations highlight the need for ongoing critical reflection throughout the EDVF process.

\begin{figure}[h]
    \centering
    \includegraphics[alt={Bar chart comparing Event A and Event B usability scores across engagement, interest, efficiency, and related metrics.},width=.9\columnwidth]{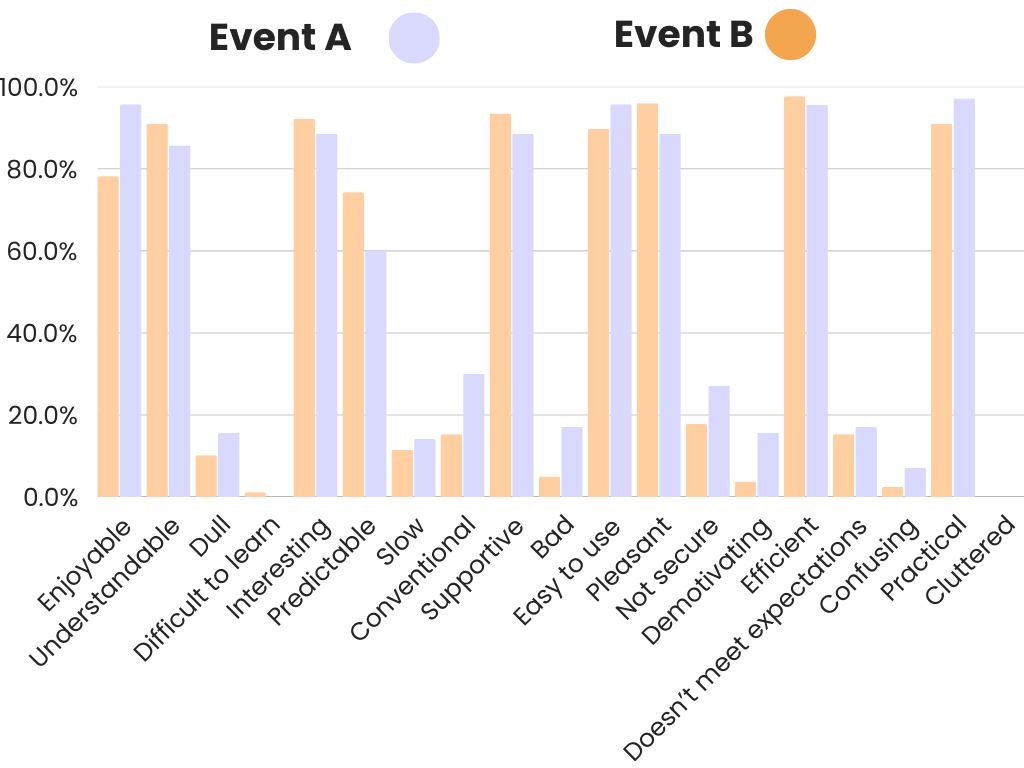}
    \vspace{-2mm}
    \caption{Results from the Usability Experience Questionnaire at Event A (Betsi Cadwaladr Hospital – Women’s Health) and Event B (Denbighshire Annual Dementia Conference), showing positive feedback on engagement, interest, and key usability factors.}
    \vspace{-3mm}
    \label{fig:ResultsEventAB}
\end{figure}

\section{Evaluation}
\label{SEC:evaluation}

To assess the \textbf{Empathy-Driven Visualisation Framework (EDVF)}, we conducted a user study of the tool created using the framework, using the UEQ at two healthcare events: the \textit{Betsi Cadwaladr Women's Health Event} and the \textit{Denbighshire Annual Dementia Care Conference}. Our evaluation followed the Design Study Methodology \cite{seldmair2012}, including briefing and consent, hands-on interaction with the prototype, and structured feedback collection through the UEQ~\cite{Laugwitz2008}, open-ended responses, and observational notes. The application (\cref{fig:caseStudy}) was developed through a design study informed by lived experience in dementia care and consultation with clinical staff. The interface prioritised rapid access to critical patient information through iconography and contextual overlays, supporting efficient yet sensitive care delivery. Across both events, 32 participants, including clinicians, care volunteers, and individuals with dementia, interacted with the prototype. Quantitative results demonstrated strong engagement and usability scores (\cref{fig:ResultsEventAB}), while qualitative feedback underscored the importance of empathetic data representation. Notably, Carol, a participant living with dementia, shared how simple changes in information presentation can dramatically influence her comfort in clinical settings, highlighting the role of visual analytics in ensuring dignity and trust in care.

\subsection{Results and Analysis}

\begin{figure}[t]
    \centering
    \includegraphics[alt={Diagram showing key themes from participant feedback, with confidence as the largest category, followed by other factors, and efficiency represented by the smallest segment.},width=0.9\columnwidth]{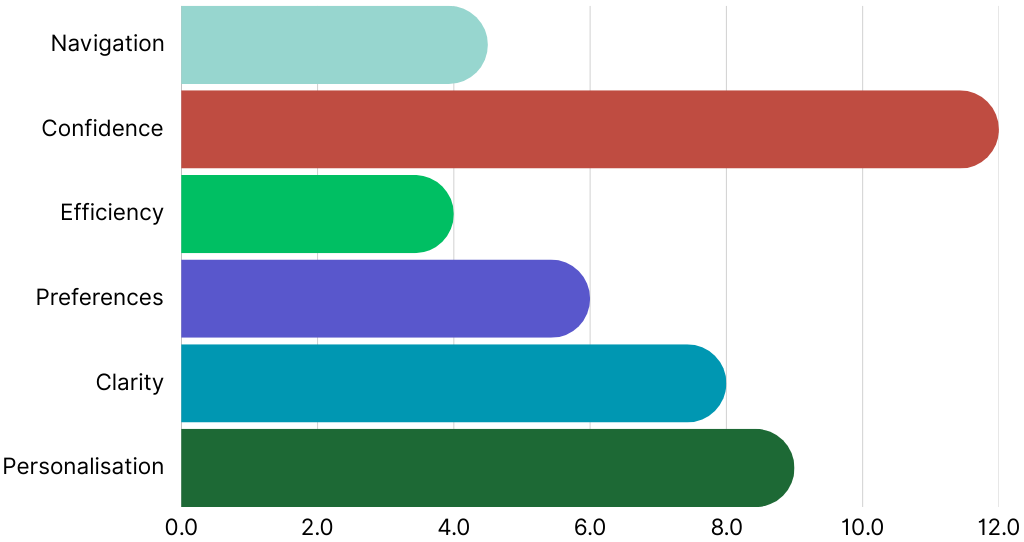}
    \caption{Thematic analysis of participant feedback. Confidence, particularly in system outputs, was most frequently cited. Efficiency received the fewest mentions, indicating an area for future emphasis.}
    \label{fig:ResultsThematic}
    \vspace{-2mm}
\end{figure}

The evaluation results demonstrate the effectiveness of the Empathy-Driven Visualisation Framework (EDVF) across clinical and care contexts. As shown in \cref{fig:ResultsEventAB} and \cref{fig:ResultsThematic}, the User Experience Questionnaire (UEQ) responses were highly favourable, with top scores for \textit{Efficient} (6.78, 96.65\%), \textit{Practical} (6.63, 94.05\%), \textit{Easy to use} (6.54, 92.70\%), and \textit{Pleasant} (6.48, 92.25\%). Negative attributes such as \textit{Cluttered} (0.50) and \textit{Confusing} (0.83) received negligible ratings, supporting the system’s clarity and cognitive accessibility.

Qualitative feedback reinforced these findings, offering user-centred insights that informed iterative design refinement. For instance, a clinician suggested that red should be avoided for wearable indicators as it's associated with allergies, recommending purple instead due to its strong association with dementia care. This kind of contextual awareness illustrates the critical value of engaging domain experts during design validation. In total, 32 participants, including clinicians, care staff, and individuals with lived experience of dementia contributed to the study. High ratings for \textit{Supportive} (91.00\%) and \textit{Understandable} (88.35\%) echo McCabe’s observation that \textit{``effective dementia tools must balance technical functionality with emotional resonance''}~\cite{mccabe2016personcentered}. 

\begin{figure}[t]
    \centering
    \includegraphics[alt={Laptop screen displaying the prototype app’s patient profile interface, featuring soft colours, icons, and structured sections for empathetic presentation.},width=.9\columnwidth]{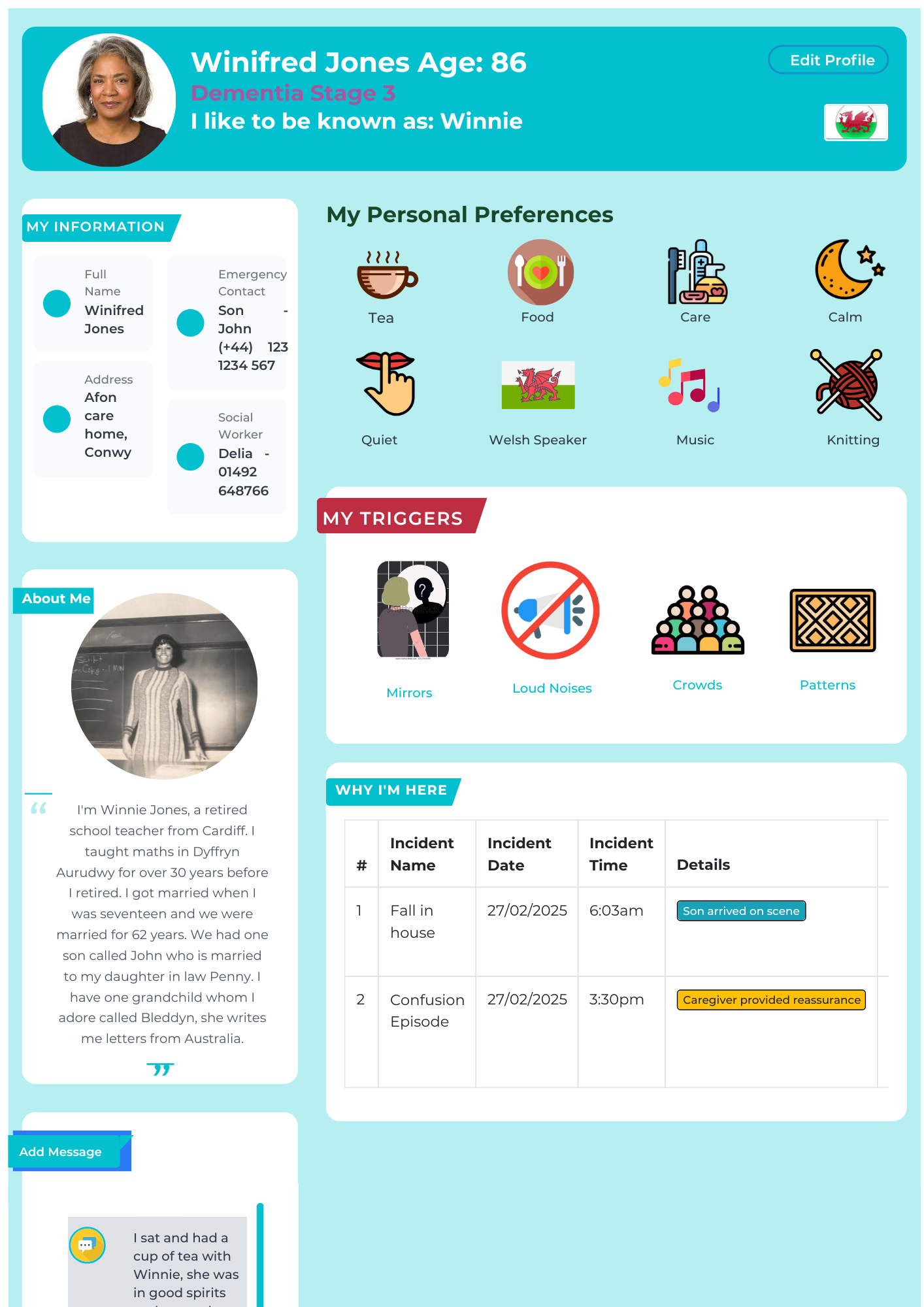}
    \caption{A screenshot of the prototype app interface as it was being used on a laptop, showing its empathetic design having used the EDVF.}
    \label{fig:caseStudy}
\end{figure}

To operationalise the EDVF in practice, we developed a prototype as a proof of concept (\cref{fig:caseStudy}), demonstrating how empathy could be embedded into the visual representation of patient data. The prototype moves away from traditional dashboard paradigms and instead presents information about a fictional dementia patient using our three core design principles: Hierarchical Salience-Driven Design, Task-Optimised Iconography, and First-Person Narrative Integration. Usability testing, aligned with established methodologies~\cite{nielsen1994usability, brooke2013sus} and the UEQ framework~\cite{Laugwitz2008}, affirmed its effectiveness: 96.65\% of participants rated the system as efficient, while only 4\% found it confusing. The EDVF’s emphasis on \textit{immersion}, defined here as deep contextual engagement with the dementia care environment, enabled the translation of complex clinical data into intuitive and emotionally resonant visual narratives. Drawing on Gerdes’ conceptual framework for empathy evaluation~\cite{gerdes2010conceptualising}, we integrated observational insights, discussion, and text analysis to capture a multi-faceted understanding of empathic impact. This aligns with calls in the visualisation community to incorporate narrative and emotional design elements~\cite{wonyounghumanevisualai, segel2010narrative}, advancing efforts to visualise the human element in patient data. To further investigate user sentiment, we conducted a thematic analysis on the qualitative data using a top-down keyword approach to classify responses into six semantic categories: Confidence, Personalisation, Clarity, Preferences, Navigation, and Efficiency. Themes were visualised using a colour-coded bar chart, with frequency counts determining bar lengths.

\cref{fig:ResultsThematic} highlights Confidence (12 mentions) as the most dominant theme, underscoring trust in the system’s interpretability and alignment with clinical workflows. Personalisation (9 mentions) and Clarity (8 mentions) also featured strongly, supporting the framework’s adaptability and ease of use. Efficiency (4 mentions), while less prominent, suggests scope for enhancing perceptions of time-saving potential. Collectively, the combined results validate the potential use of the EDVF as an guiding framework for creating visualisation tools in healthcare.

\section{Discussion and Future Work}

This work introduces the Empathy-Driven Visualisation Framework (EDVF), a novel approach to embedding empathy into healthcare visualisation. Our prototype demonstrated strong user approval—96.65\% rated it efficient and only 4\% found it confusing. By translating patient data into emotionally resonant, personalised visual narratives, EDVF supports more meaningful clinician-patient communication and enhances quality of care.

Medical data is inherently complex and often highly technical, requiring time to interpret and explain. In dementia care, this complexity poses unique challenges: patients are frequently unable to fully comprehend their situation, leaving family members and loved ones to receive and process the information. At the same time, nurses and care staff face significant time pressures, making rapid and accurate communication essential. In such contexts, visualisations must not only be empathetic, clear, and sensitive, but also carefully designed to minimise the risk of misinterpretation—particularly under stress. Plots and charts can be misleading if overloaded with detail or poorly contextualised, and colour choices can carry different meanings for different viewers, potentially causing confusion. Effective visual design in dementia care therefore requires attention to accessibility, cultural interpretation, and cognitive load, ensuring that critical information is both accurate and immediately understandable.

While there is growing interest in feminist approaches, inclusivity, diversity in data visualisation, and human–computer interaction work that engages with empathy, there remains relatively little research specifically focused on empathetic visualisation. The concept itself is still not clearly defined, and there is limited consensus on how it can be effectively achieved in practice. The work presented here begins to address this gap, offering a framework that can help others navigate these challenges and encourage both practitioners and researchers to think critically about the role of empathy in visual communication. In particular, adapting and extending the ``This Is Me'' paper tool to suit technological contexts and the needs of nurses working with digital systems represents a step toward more compassionate, human-centred design in dementia care.

We are preparing a real-world deployment of a fully functional \textit{``About Me''} application in dementia care wards at Glan Clwyd and Llandudno hospitals. Concurrently, we are initiating a new collaboration with NHS Wales to visualise \textit{The Women's Health Plan for Wales}, a 10-year vision to transform women's healthcare services, applying EDVF principles to address gender-specific health disparities. These complementary projects will serve as foundational implementations demonstrating EDVF's potential to humanise diverse healthcare contexts. With support from the local dementia care community~\cite{DVSCwebsite, AbergeleDementia}, we aim to assess the framework's impact on personalised health management at scale. Future work will incorporate explainable AI techniques, such as longitudinal preference modelling, to detect and visualise subtle shifts in patient needs, making care more adaptive and person-centred. Our findings contribute to the growing discourse on patient-generated health data, personalised care, empathetic visualisation and emotionally responsive interfaces in clinical VA systems. Ultimately, EDVF represents a step toward more compassionate, interpretable, and patient-aligned healthcare technologies.

\acknowledgments{
The authors would like to express their gratitude to those who have supported this research. Special thanks to Llandudno Hospital, and especially Kathryn Williams, the dementia community, BCUHB, and the study participants including Carol for their invaluable assistance. The evaluations were approved under the following ethics clearance number: CSE-2025-0721. The work was supported by the UKRI AIMLAC CDT EP/S023992/1.
}

\bibliographystyle{abbrv-doi-hyperref}

\bibliography{template}
\end{document}